\documentclass[12pt]{article}
\usepackage{latexsym,graphicx,multirow}
\usepackage{amssymb}
\usepackage{amsmath}
\usepackage{amscd}
\usepackage{amsthm}
\usepackage[left=2cm,top=2.5cm,right=2.5cm,bottom=1.5cm]{geometry}
\usepackage{hyperref}
\usepackage{color}

\usepackage{epstopdf}
\usepackage{cite}
\usepackage{float}
\usepackage[utf8]{inputenc}

\begin{document}

    \begin{center}
        \large{\bf{New Tsallis Holographic Dark Energy}} \\
        \vspace{10mm}
   \normalsize{ Bramha Dutta Pandey$^{a,1}$, Suresh Kumar P$^{b,2}$, Pankaj$^{c,3}$, Umesh Kumar Sharma$^{d,4}$ } \\
    \vspace{5mm}
    \normalsize{$^{a,\:b,\:d}$ Department of Mathematics, Institute of Applied Sciences and Humanities, GLA University
        Mathura-281406, Uttar Pradesh, India}\\
    \vspace{2mm}
    \normalsize{$^{a,\:b}$ IT Department (Math Section), University of Technology and Applied Sciences-Salalah, Oman}\\
    \vspace{2mm}
    \normalsize{$^c$ IT Department (Math Section), University of Technology and Applied Sciences-HCT, Muscat, Oman}\\
    \vspace{2mm}

    $^1$E-mail: bdpandey05@gmail.com\\
    $^2$E-mail: sureshharinirahav@gmail.com\\
    $^3$E-mail: pankaj.fellow@yahoo.co.in \\
    $^4$E-mail: sharma.umesh@gla.ac.in \\
    \vspace{10mm}

\end{center}

\begin{abstract}
Tsallis entropy is a generalization of the Boltzmann-Gibbs entropy in statistical theory which uses a parameter $\delta$ to measure the deviation from the standard scenario quantitatively. Using concepts of Tsallis entropy and future event horizon, we construct a new Tsallis holographic dark energy model. The parameters $c$ and $\delta$ will be used to characterize various aspects of the model. Analytical expressions for various cosmological parameters such as the differential equation describing the evolution of the effective dark energy density parameter, the equation of state parameter and the deceleration parameter are obtained. The equation of state parameter for the current model exhibits the pure quintessence behaviour for $c>1$, quintom behaviour for $c<1$ whereas the $\Lambda$CDM model is recovered for $c=1$. To analyze the thermal history of the universe, we obtained the expression for the deceleration parameter and found that for $z \approx 0.6$, the phase transits from deceleration to acceleration. \\ 
\smallskip


\end{abstract}

\section{\textbf{Introduction}}
Based on the observations carried out by Reiss and Perlmutter \cite{Riess98,Perl99} the current universe is in an accelerated expansion phase. The reason behind it can be considered as the existence of cosmological constant $\Lambda$. But the dynamic behaviour of $\Lambda$ and computing the value of it in a quantum field theoretic way opens the paths of  extended scenarios. One of which is to keep general relativity based on gravitation theory while considering new, exotic matter, which explains the dark energy (DE) concept \cite{Copeland06,Cai10,Bamba12}. The other is to extend the theory of gravity whose special case is general relativity with extended degree of freedom to get the explanation about the accelerated universe \cite{Nojiri10,Capozziello11,Cai16,CANTATA21}. \\

Using the holographic Principle (HP), one intriguing possibility for explaining the genesis and nature of dark energy can be obtained at a cosmological framework \cite{Hooft93,Susskind95,Bousso02}. Gerard 't Hooft \cite{Hooft93} presented the well-known concept of holographic principle based on investigations of black hole thermodynamics \cite{Bekenstein73,Hawking75}. This asserts that a hologram corresponding to a theory on the volume's border can be used to represent the whole information contained in a spatial volume. HP was used to solve the DE problem by proposing the concept of holographic dark energy (HDE) model \cite{Li04}. According to HDE model, on the universe's edge, the reduced Plank's mass $M_p\equiv\dfrac{1}{\sqrt{8\pi G}}$, where $G$ denotes the  universal gravitational constant of Newton and universe's  future event horizon $L$ \cite{Li04} are the physical quantities on which the dark energy's density $\rho_d$ depends . The DE model equipped with HP (HDE) supports the current cosmological observations \cite{Li04,SWang17,Horvat04,Pavon05,Nojiri06,BWang06,Setare09Saridakis} and extensively studied \cite{Cai07,Jawad16,Pasqua16,Pourhassan18,Nojiri17}. Also observational data are in agreement with the concept of HDE \cite{Zhang05,Li09,Feng07,Zhang09,Lu10}.\\

Due to the long-range nature of gravity and the unpredictable structure of spacetime, many extended entropy formalisms have been employed to investigate gravitational and cosmic phenomena. In order to study gravitational and cosmic systems through the concepts of  generalized statistical mechanics, the Tsallis's entropy \cite{Tsallis12,Tavayef:2018xwx} plays a central role. Kaniadakis, on the other hand, presented generalized Boltzmann-Gibbs entropy through single parameter, known as Kaniadakis entropy \cite{Kaniadakis02,Kaniadakis05} and studied by Niki and Sharma \cite{Drepanou21, Sharma21} using the concepts of future event horizon and apparent horizon respectively. This is the consequence of a unified and self-consistent relativistic statistical theory that retains the core properties of normal statistical theory. The usual Maxwell-Boltzmann statistics is continuously deformed by a single parameter leading to the extended statistical theory, whose limiting case is the standard statistical theory.
In this manuscript we will apply the Tsallis entropy concept to formulate  new Tsallis holographic dark energy (NTHDE) by considering the future event horizon as an IR cut-off and investigate its cosmic implications.  The study of NTHDE carried out by \cite{Moradpour20,Jawad21, Nisha21} is based on the consideration of Hubble horizon as an IR cut-off which could not recover the standard HDE model for which Tsallis entropy should become the standard entropy but it is not. Large parameter values could represent the universe's evolution. Such a consideration results in more deviation from the standard entropy. This difference is due to the Hubble horizon acting as an IR cut-off. As a result, in this paper, we develop a consistent formulation of NTHDE to get a well-defined extension of conventional HDE, which is the limiting case when the Tsallis entropy becomes the conventional Bekenstein Hawking entropy. \\

In section 2, by formulating the NTHDE expression, the differential equation for specific DE density parameter $\Omega_d$, expressions for deceleration parameter and equation of state (EoS) parameter are obtained analytically. Section 3 is devoted to studying cosmological behaviour. In section 4, a discussion on the obtained results is carried out with a concluding summary.

\section{\textbf{New Tsallis Holographic Dark Energy}}
NTHDE formulation will be established here. The DE density $\rho_d$, the entropy $S$ of black hole with radius $L$ and the largest theory's distance $L$ connected by the relation $\rho_d L^4\leq S$ is the key idea for HDE formulation \cite{Li04,SWang17}. For usual Bekenstein-Hawking entropy $S_{BH}\propto (4G)^{-1}A=\pi G^{-1}L^2$ with Newton's gravitational constant $G$. The standard HDE $\rho_d=3c^2M_p^2L^{-2}$  with model parameter $c$ is the saturation of the above inequality. As a result, a modified HDE model is obtained by modifying the entropy.\\

If $k_B=1$ and a distribution has $W$ states with Gibbs and Shannon entropies, then the expression for each state is same and given by 

\begin{equation}
S=-\sum_{i=1}^W P_i \:\mathrm{ln}(P_i). \label{P2_1.1}
\end{equation}
The Von-Neumann entropy or the quantum mechanical equivalent of (\ref{P2_1.1}) is
\begin{equation}
S=-\mathrm{Tr} [\rho \: \mathrm{ln}(\rho)].   \label{P2_1.2}
\end{equation}
For classical systems, (\ref{P2_1.2}) supports Boltzmann's proposal in phase space with state density $\rho$. The Bekenstein-Hawking entropy $\left(\equiv S_{BH}=\dfrac{A}{4}\right)$ is obtained by applying (\ref{P2_1.2}) to a pure gravitational system where $A$ is system's area \cite{Srednicki93}. By assumption that the degrees of freedom are dispersed on the horizon where no particular priority for each other is specified \cite{Das07,Pavon20}, all $P_i$'s are equal and $P_i=\dfrac{1}{W}$. Both (\ref{P2_1.1})and (\ref{P2_1.2}) implies the Boltzmann's entropy ($S=\mathrm{ln}(W)$) and hence we get the expression for horizon entropy \cite{Abreu18}
\begin{equation}
S_{BH}=\dfrac{A}{4}=\mathrm{ln}(W)\rightarrow W=\mathrm{e}^{\left(\dfrac{A}{4}\right)}.  \label{P2_1.3}
\end{equation}
The Tsallis entropy is defined by \cite{Masi05}
\begin{equation}
S_n^T=\dfrac{1}{1-n}\sum_{i=1}^W (P_i^n-P_i)=\dfrac{W^{1-n}-1}{1-n},   \label{P2_1.4}
\end{equation}
where $P_i=\dfrac{1}{W}$, $n$ is an unknown parameter (non-extensive) and as $n\rightarrow 1, S_n^T\rightarrow S$. The parameter $n$ may also have its roots in quantum features of gravity. Using (\ref{P2_1.3}),
(\ref{P2_1.4}) and $1-n=\delta$ we get
\begin{eqnarray}
S_n^T=\dfrac{1}{1-n}\left[ \mathrm{e}^{(1-n)S_{BH}}-1\right], \notag \\
S_\delta^T=\dfrac{2\:\mathrm{e}^{\left(\dfrac{\delta \:S_{BH}}{2}\right)}}{\delta} \mathrm{sinh} \left( \dfrac{\delta\:S_{BH}}{2} \right).  \label{P2_1.5}
\end{eqnarray}
As $\delta \rightarrow 0$ the standard Bekenstein-Hawking entropy is recovered. As expected the usual Bekenstein-Hawking entropy is obtained as a limiting case of Tsallis entropy and hence $\delta\ll1$ i.e. $\delta\in (-1,1)$. The equation (\ref{P2_1.5}) in its expanded and truncated form is given by 
\begin{equation}
S_\delta^T=S_{BH}+\dfrac{\delta \: S_{BH}^2}{2}+\dfrac{\delta^2 \: S_{BH}^3}{6}+\mathcal{O}(\delta^3) \label{P2_1.6}
\end{equation}
Clearly, the first term of (\ref{P2_1.6}) is the standard entropy. Using (\ref{P2_1.6}) and  $\rho_d L^4\leq S$ we get
\begin{equation}
\rho_d=\dfrac{3 c^2 M_p^2}{L^2}+\dfrac{3 c_1^2 \delta M_p^4}{2}+\dfrac{3 c_2^2 \delta^2 M_p^6 L^2}{6} \label{P2_1.7}
\end{equation}
where $c,\:c_1,\: c_2$ are constants. For $\delta=0,$ the equation (\ref{P2_1.7}) leads to standard HDE, i.e $\rho_d= \dfrac{3 c^2 M_p^2}{L^2}$. By letting $\dfrac{3 c_1^2 \delta}{2}=\delta_1$ and $\dfrac{3 c_2^2 \delta^2}{6}=\delta_2^2$, equation (\ref{P2_1.7}) can be rewritten by absorbing $c_1$ and $c_2$ in $\delta$ as 
\begin{equation}
\rho_d=\dfrac{3 c^2 M_p^2}{L^2}+ \delta M_p^4 + \delta^2 M_p^6 L^2 \label{P2_1.8}
\end{equation}
By considering the geometry of Friedmann-Robertson-Walker (FRW) model to be homogeneous, isotropic and flat with metric described by
\begin{equation}
\mathrm{d}s^2=\delta_{ij} \mathrm{d}x^i \mathrm{d}x^ja^2(t) -\mathrm{d}t^2 ,\label{P2_1.9}
\end{equation}
with a scaling factor $a(t)$  that varies with cosmic time. To investigate an HDE model the largest distance $L$ of the theory is needed. According to Li and Hsu \cite{Li04,Hsu04}, $L\neq H^{-1}$ is the need for an HDE model to be consistent and standard. The Hubble horizon is expressed as $H^{-1}=\dot{a}(t)^{-1} \: a(t)$. Future event horizon as offered by Li \cite{Li04} is expressed as 
\begin{equation}
r_h=a(t)\int_{t}^\infty \frac{\mathrm{d}y}{a(y)}=a(t)\int_{a(t)}^\infty \frac{\mathrm{d}a}{H\left[a(y)\right]^2} . \label{P2_1.10}
\end{equation} 
In \cite{Moradpour20,Jawad21}, $L=H^{-1}$ is considered as IR-cutoff and the parameter $\delta$ is $\mathcal{O} \left(10^3\right)$. Such a high value of $\delta$ leads to high deviation from basic Bekenstein-Hawking entropy. We want to construct NTHDE consistently in this paper, thus we utilize the future event horizon $r_h$ as $L$ in (\ref{P2_1.8}) and get the NTHDE density as 

\begin{equation}
\rho_d=\dfrac{3 c^2 M_p^2}{r_h^2}+ \delta M_p^4 + \delta^2 M_p^6 r_h^2 \label{P2_1.11}
\end{equation}
Friedmann's equations for a universe made up of  perfect fluids such as DE and dark matter are expressed by
\begin{eqnarray}
3 M_p^2 H^2=\rho_m+\rho_d  , \label{P2_1.12}
\\
-2 M_p^2 \dot{H}=P_m+P_d+\rho_m+\rho_d ,\label{P2_1.13}
\end{eqnarray} 
where $P_d$ represents NTHDE pressure, $\rho_m$ represents dark matter energy density, and $P_m$ represents dark matter pressure. The dark matter conservation equation is as follows:
\begin{equation}
\dot{\rho}_m+3 H (\rho_m + P_m)=0 .\label{P2_1.14}
\end{equation}
The fractional DE and dark matter density parameters are defined as 
\begin{eqnarray}
\Omega_d&=&\frac{\rho_d}{3 M_p^2 H^2} , \label{P2_1.15} 
\\
\Omega_m&=&\frac{\rho_m}{3 M_p^2 H^2}  \label{P2_1.16}
\end{eqnarray}
respectively. Using the equation (\ref{P2_1.11}) in (\ref{P2_1.15}), we get a fourth degree equation in $r_h$. By considering  $r_h$ to be positive and taking the limit $\delta \rightarrow 0$ the standard HDE $\int_{x}^\infty \frac{\mathrm{d}x}{Ha}=\dfrac{c}{H a \sqrt{\Omega_d}}$ is obtained. Hence such a value of $r_h$ is considered and expressed by
\begin{equation}
r_h=\left(\dfrac{3 H^2 \Omega_d-\delta M_p^2-\sqrt{(3 H^2 \Omega_d-\delta M_p^2)^2-12 c^2 \delta^2 M_p^4 } }{2 \delta^2 M_p^4} \right)^{\dfrac{1}{2}}.\label{P2_1.17}
\end{equation}
Using equation (\ref{P2_1.10}) and (\ref{P2_1.17}), we get 
\begin{equation}
\int_{x}^\infty \frac{\mathrm{d}x}{Ha}= \dfrac{1}{a}\left(\dfrac{3 H^2 \Omega_d-\delta M_p^2-\sqrt{(3 H^2 \Omega_d-\delta M_p^2)^2-12 c^2 \delta^2 M_p^4 } }{2 \delta^2 M_p^4} \right)^{\dfrac{1}{2}}, \label{P2_1.18}
\end{equation} 
where $a=\mathrm{e}^x$.

Now we consider the physically intriguing dust matter scenario for which the matter EoS parameter is zero. If we consider the current the matter energy density to be $\rho_{m_0}$ for current scale factor $a_0=1$, equation (\ref{P2_1.14}) gives
\begin{equation}
\rho_m=\dfrac{\rho_{m_0}}{a^3}. \label{P2_1.19}
\end{equation}
Using equation (\ref{P2_1.19}) into (\ref{P2_1.16}) we get
\begin{equation}
\Omega_m=\dfrac{\Omega_{m_0} H_0^2}{H^2a^3}, \label{P2_1.20}
\end{equation}
Here $H_0$ is called the Hubble constant. 
\\Using equation (\ref{P2_1.20}) and the Friedmann equation $\Omega_d+\Omega_m=1$ we get 
\begin{equation}
\dfrac{1}{Ha}=\dfrac{\sqrt{a (1-\Omega_d)}}{ H_0\sqrt{\Omega_{m_0}}}. \label{P2_1.21}
\end{equation}

Substituting equation (\ref{P2_1.21}) into (\ref{P2_1.18}) we get 
\begin{equation}
\int_{x}^\infty \dfrac{\sqrt{a (1-\Omega_d)}}{ H_0\sqrt{\Omega_{m_0}}} \mathrm{d}x= \dfrac{1}{a}\left(\dfrac{3 H^2 \Omega_d-\delta M_p^2-\sqrt{(3 H^2 \Omega_d-\delta M_p^2)^2-12 c^2 \delta^2 M_p^4 } }{2 \delta^2 M_p^4} \right)^{\dfrac{1}{2}}, \label{P2_1.22}
\end{equation}

Differentiating equation (\ref{P2_1.22}) with respect to `$x$' we get
\begin{equation}
\Omega_d ' = \Omega_d (1-\Omega_d)\left[ 3-\dfrac{2\left( \mathcal{I}-2 \delta^2 M_p^6 \mathcal{J} \right)}{\mathcal{I}+\delta M_p^4} \left\lbrace 1-\sqrt{3} \left( \dfrac{M_p^2 \Omega_d}{\left( \mathcal{I}+\delta M_p^4 \right)\mathcal{J}}  \right)^{\dfrac{1}{2}}\right\rbrace\right], \label{P2_1.23}
\end{equation}
where
\begin{eqnarray}
\mathcal{I}&=&\dfrac{3\: \mathrm{e}^{-3x} H_0^2 M_p^2 \Omega_{m_0} \Omega_d }{1-\Omega_d}-\delta M_p^4, \notag 
\\
\mathcal{J}&=&\dfrac{\mathcal{I}-\sqrt{\mathcal{I}^2-12 c^2 \delta^2 M_p^8}}{2 \: \delta^2 M_p^6}. \notag
\end{eqnarray}

For flat spatial geometry and dust matter, the differential equation (\ref{P2_1.23}) describes the evolution of NTHDE. As a limit on considering $\delta \rightarrow 0$ we get $\mathcal{J}=\dfrac{3 c^2 }{\mathcal{I}}$, which implies (\ref{P2_1.23}) to recover the differential equation of standard HDE \cite{Zhang07}, i.e. $\Omega_d ' = \Omega_d (1-\Omega_d) \left(1+\dfrac{2}{c}\sqrt{\Omega_d}\right)$ and can be solved analytically. 

Now we will consider the EoS parameter for NTHDE defined by $w_d=\dfrac{P_d}{\rho_d}$. As the matter sector is conserved. The equation (\ref{P2_1.14}) and the Friedmann equations (\ref{P2_1.12}), (\ref{P2_1.13}) implies the DE sector to be conserved, i.e.
\begin{equation}
\dot{\rho}_d+3 H \rho_d (1+w_d)=0. \label{P2_1.24}
\end{equation}
Differentiating (\ref{P2_1.11}) w.r.t. `$t$' results
\begin{equation}
\dot{\rho}_d=-\dfrac{2\: M_p^2 \left( 3 c^2 -\delta^2 M_p^4 r_h^4 \right)\dot{r_h}}{r_h^3} .\label{P2_1.25}
\end{equation}
From equation (\ref{P2_1.10}) we get 
\begin{equation}
\dot{r}_h=H r_h-1.\label{P2_1.26}
\end{equation}
Using equations (\ref{P2_1.25}) and (\ref{P2_1.26}) we get the expression for $r_h$ in terms of $\rho_d$ given by
\begin{equation}
r_h=\left[ \dfrac{\rho_d - \delta M_p^4 -\sqrt{\left(\rho_d-\delta M_p^4 \right)^2-12 c^2 \delta^2 M_p^8}}{2 \delta^2 M_p^6} \right]^{\dfrac{1}{2}}. \label{P2_1.27}
\end{equation}
Using equations (\ref{P2_1.15}), (\ref{P2_1.21}) and (\ref{P2_1.25}) to (\ref{P2_1.27}), we get
\begin{equation}
w_d = -1 -2 \left( \dfrac{M_p^6 \: \Omega_d}{3\left(\mathcal{I}+\delta M_p^4 \right)^3}\right)^{\dfrac{1}{2}} \left(\dfrac{\delta^2 M_p^4 \mathcal{J}^2-3c^2}{\mathcal{J}^{\dfrac{3}{2}}} \right) \left[-1+\left(\dfrac{\left( \mathcal{I}+\delta M_p^4\right)\mathcal{J}}{3 M_p^2 \Omega_d} \right)^{\dfrac{1}{2}} \right],  \label{P2_1.28}
\end{equation}
Clearly the standard HDE is recovered by letting $\delta \rightarrow 0$ i.e. as $\delta \rightarrow 0, \: w_d \rightarrow \dfrac{-1}{3}-\dfrac{2 \sqrt{\Omega_d}}{3 c}$. In general, we can highlight that $w_d$ can behave either like quintessence or quintom which shows the richness of the current model. 

The parameter describing deceleration behaviour can be expressed as 
\begin{equation}
q=-\dfrac{\dot{H}}{H^2}-1=\dfrac{3\:w_d \Omega_d+1}{2}.  \label{P2_1.29}
\end{equation}

\section{\textbf{Cosmological evolution of NTHDE}}

In section 2, we derived the differential equation describing the evolutionary behaviour of NTHDE density parameter, corresponding expressions for EoS  and deceleration parameters. Now we will discuss the detailed cosmological behaviour for results obtained in the previous section. The numerical solution for the differential equation (\ref{P2_1.23}) reflects various evolutionary features of $\Omega_d$ for redshift $z$ by the transformation $x=\mathrm{ln} \left(\dfrac{1}{1+z} \right)$ with initial condition $\Omega_d (x=0)=\Omega_d[0]\approx 0.7$. And hence by virtue of Friedmann equation $\Omega_{m_0}\approx 0.3$. 
\begin{figure}
[H]
\includegraphics[scale=0.7]{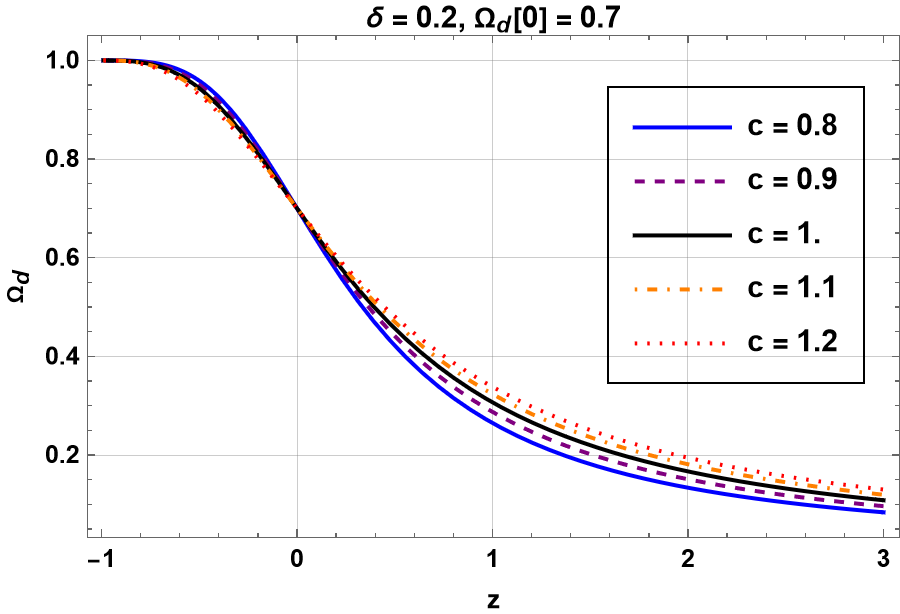}
\centering
\label{P2_z_Om_c} 
\caption
{
	\begin{small}
		NTHDE density parameter $\Omega_d$ with $\delta=0.2$ and $c=0.8$ to $1.2$ is plotted w.r.t. redshift $z$ by considering  $\Omega_d[z=0] \approx 0.7, \: M_p^2=1 $.
	\end{small}
}
\end{figure}
\begin{figure}
[H]
\includegraphics[scale=0.7]{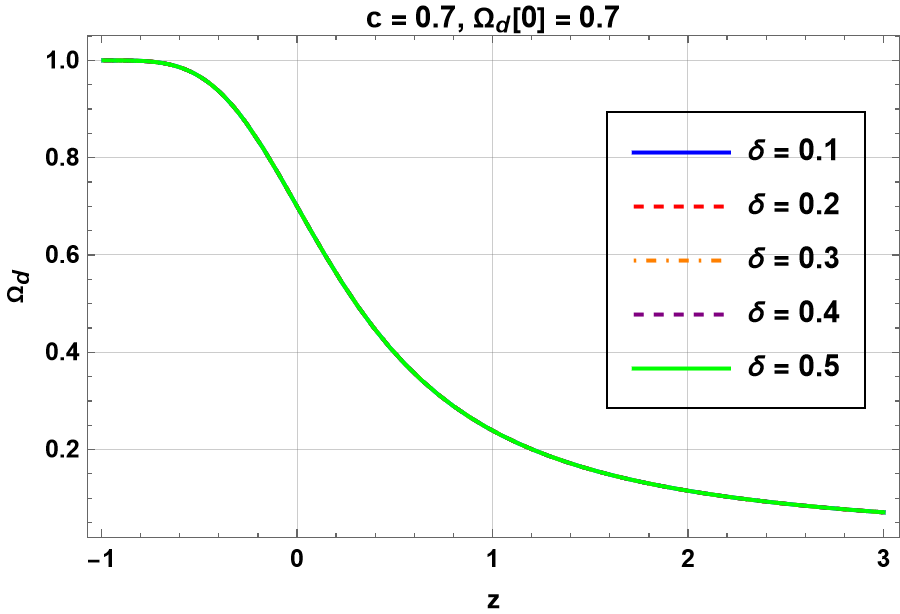}
\centering
\label{P2_z_Om_delta} 
\caption
{
	\begin{small}
		NTHDE density parameter $\Omega_d$ with $c=0.7$ and $\delta=0$ to $0.4$ is plotted w.r.t. redshift $z$ by considering  $\Omega_d[z=0] \approx 0.7, \: M_p^2=1 $.
	\end{small}
}
\end{figure}
The Fig. 1 and 2 shows the DE density parameter plots against the redshift $z$. In Fig. 1, we have considered $\delta=0.2$ fixed with varying $c$ values. Fig. 2 is plotted by considering $c=0.7$ fixed and varying $\delta$. As we can see from both the graphs, the current model may give the universe's needed thermal history, i.e. in the past matter dominated, current domination of $70\%$ by DE and in future fully dominated by DE only. 
\begin{figure}
[H]
\includegraphics[scale=0.7]{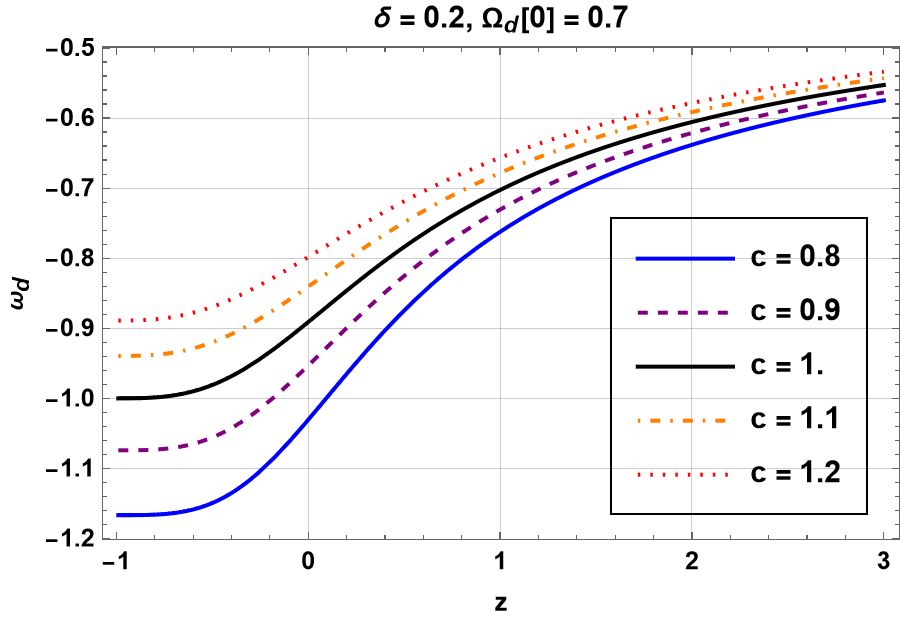}
\centering
\label{P2_z_w_c} 
\caption
{
	\begin{small}
		The evolution of EoS parameter $w_d$ of NTHDE with $\delta=0.2$ and $c=0.8$ to $1.2$ is plotted w.r.t. redshift $z$ by considering  $\Omega_d[z=0] \approx 0.7, \: M_p^2=1 $.
	\end{small}
}
\end{figure}
\begin{figure}
[H]
\includegraphics[scale=0.7]{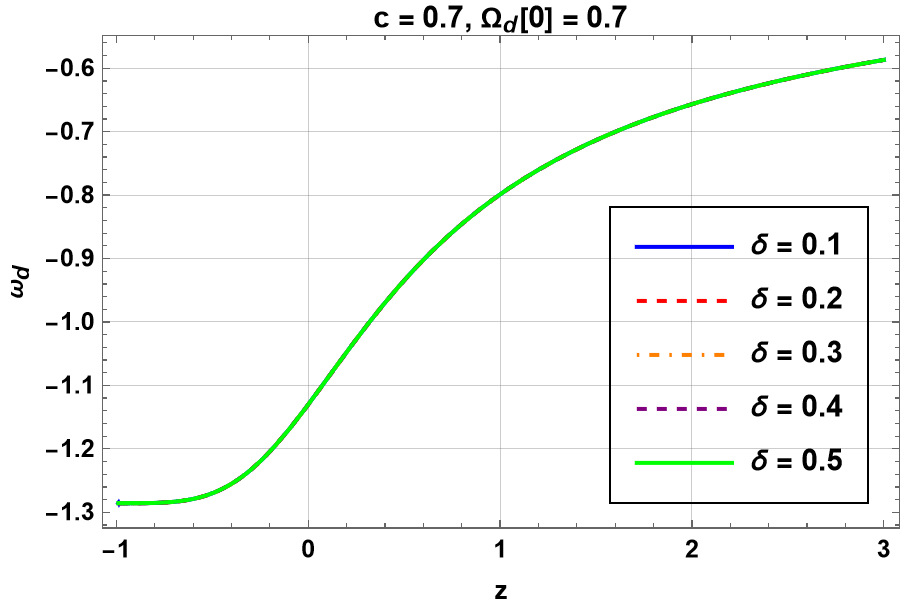}
\centering
\label{P2_z_w_delta} 
\caption
{
	\begin{small}
		The evolution of EoS parameter $w_d$ of NTHDE with $c=0.7$ and $\delta=0$ to $0.4$ is plotted w.r.t. redshift $z$ by considering  $\Omega_d[z=0] \approx 0.7, \: M_p^2=1 $.
	\end{small}
}
\end{figure}
Fig. 3 and 4 represent EoS parameters for the NTHDE model. Which shows that the current value of $w_d$ resides in the vicinity of $-1$, which is consistent with the observational data. Now we'll look at how the model parameters $\delta$ and $c$ affect the DE's EoS parameter $w_d$. In Fig. 3 we have plotted $w_d$ for  $\delta=0.2$ and different $c$ values. As can be seen, for $c<1$ values, $w_d$ always enters the phantom regime in the far future. While $c>1$ completely lies in the quintessence region. $c=1$ corresponds to the $\Lambda$CDM model. In addition, we show $w_d$ for constant $c=0.7$ and different $\delta$ values in Fig. 4. We have an intriguing pattern here with rising $\delta$, $w_d$ remains almost the same at times around the current ones. To get the far future value of $w_d$, i.e. for $z \rightarrow -1$ equation (\ref{P2_1.28}) indicates the combined dependence on $\delta$ and $c$. In conclusion NTHDE leads to some fascinating cosmic phenomenology where $w_d$ shows behaviours like quintessence, or like quintom. 
\begin{figure}
[H]
\includegraphics[scale=0.7]{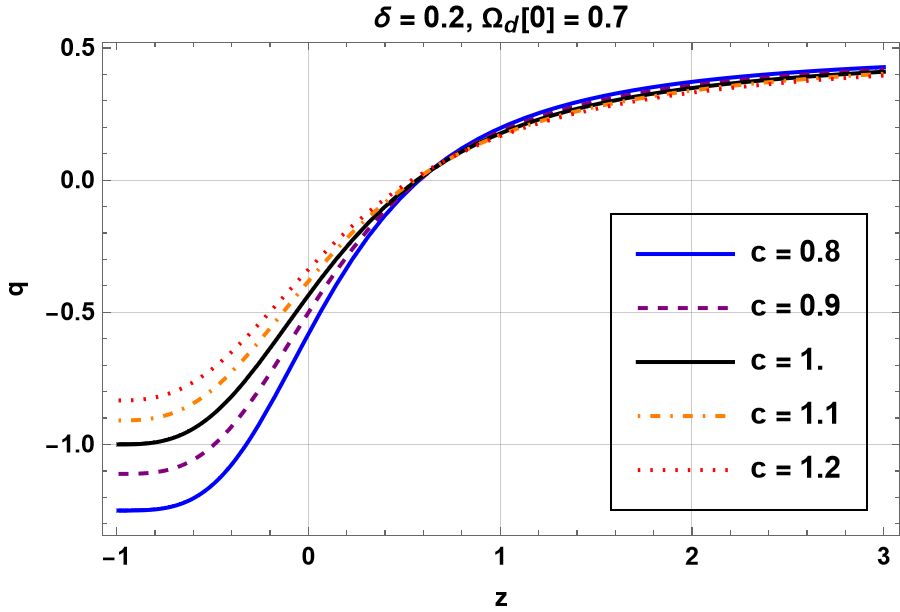}
\centering
\label{P2_z_q_c} 
\caption
{
	\begin{small}
		The deceleration parameter $q$ with $\delta=0.2$ and $c=0.8$ to $1.2$ is plotted w.r.t. redshift $z$ by considering  $\Omega_d[z=0] \approx 0.7, \: M_p^2=1 $.
	\end{small}
}
\end{figure}
\begin{figure}
[H]
\includegraphics[scale=0.7]{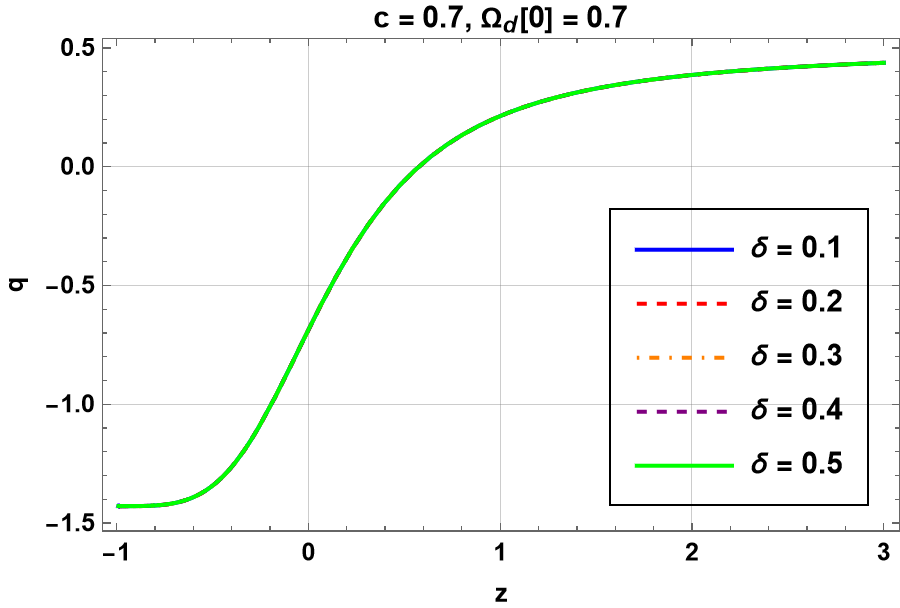}
\centering
\label{P2_z_q_delta} 
\caption
{
	\begin{small}
		The deceleration parameter $q$  with $c=0.7$ and $\delta=0$ to $0.4$ is plotted w.r.t. redshift $z$ by considering  $\Omega_d[z=0] \approx 0.7, \: M_p^2=1 $.
	\end{small}
}
\end{figure}

Fig. 5 and 6 describe the deceleration parameter $q$ behaviour against $z$. Fig. 5 is plotted by fixing $\delta$ to be $0.2$ and varying $c$ values. While Fig. 6 is based on varying $\delta$ and fixed $c=0.7$. It confirms the universe to enter an accelerated phase for $z\approx0.6$.
Which is in full agreement with the observational data supported by \cite{Riess98,Perl99}. 

\begin{figure}
[H]
\includegraphics[scale=0.7]{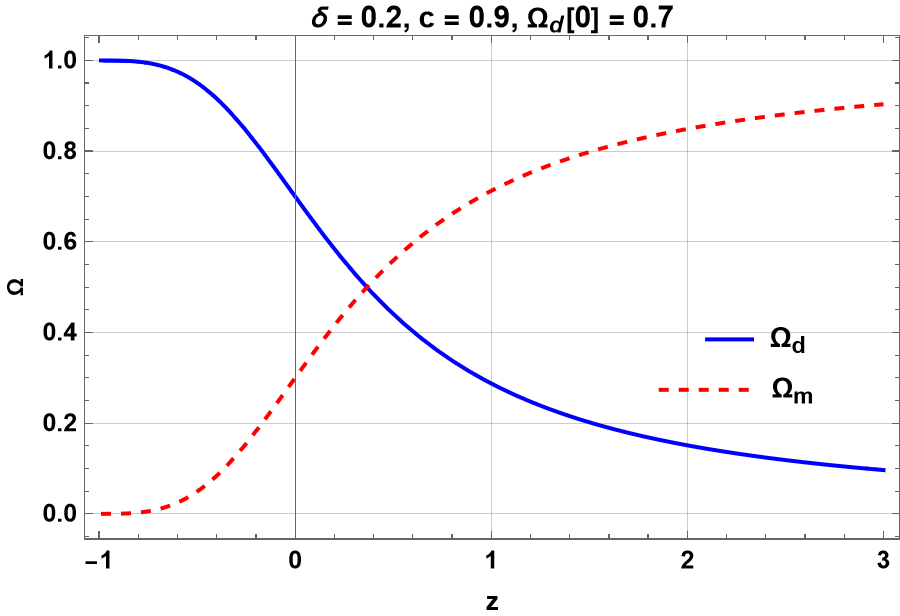}
\centering
\label{P2_z_w_Om_q} 
\caption
{
	\begin{small}
		The NTHDE density parameter $\Omega_d$ and dark matter density parameter $\Omega_m$ is depicted. The graph is plotted against the redshift $z$ by considering $\delta=0.2, \: c=0.9, \: \Omega_d[z=0] \approx 0.7, \: M_p^2=1 $. Where the present time corresponds to $z=0$.
	\end{small}
}
\end{figure}

\section{\textbf{Conclusive remarks}}
In the present work we formulated the HDE model in which Tsallis entropy, a one-parameter generalization of Boltzmann-Gibbs entropy, is used.
Such a concept is derived from a consistent relativistic statistical theory. A parameter $\delta$ is used to distinguish deviations from conventional entropy expressions. The consistent NTHDE model is obtained by applying IR cutoff in terms of future event horizon and the Tsallis entropy, to the standard HDE model. The parameter $\delta$ is responsible for such an extension with usual HDE  as a limiting case $\delta \rightarrow 0$. We derived the differential equation to describe the evolutionary behaviour of dark energy density parameter $\Omega_d$ which investigates possible cosmic applicability of NTHDE. On considering today's universe to be dominated $70\%$ by DE, Fig. 7, clearly indicates the full domination of the universe by DE in the far future. In addition, the analytical formulations of the deceleration parameter and the EoS parameter are obtained. As per the observation from NTHDE's EoS parameter, the parameters $c$ and $\delta$ describe the diversified behaviour of the model i.e. pure quintessence for $c>1$, quintom for $c<1$ (in near or far future) and $\Lambda$CDM for $c=1$. The trend shown by the deceleration parameter $q$ for the model, possesses interesting cosmological descriptions such as the universe's thermal history from dark matter to DE. The transition from decelerated to accelerated phase happens at $z \approx 0.6$. Finally, because of consistent formulation and versatile behaviour, the NTHDE leads to standard HDE as a limiting case, which is the biggest advantage of the model. In order for the NTHDE to be a successful alternative to describe the DE, the model parameters must be constrained. Such constraints can be obtained using the observational data from the Hubble parameter, CMB, BAO, and SNIa. The phase-space can be analyzed to understand the global dynamics of the DE. 


\begin{thebibliography}{99}
	
\bibitem{Riess98}
A.~G.~Riess \textit{et al.} [Supernova Search Team],
``Observational evidence from supernovae for an accelerating universe and a cosmological constant'',
Astron. J. \textbf{116}, 1009-1038 (1998)
doi:10.1086/300499
[arXiv:astro-ph/9805201 [astro-ph]].

\bibitem{Perl99}
S.~Perlmutter \textit{et al.} [Supernova Cosmology Project],
``Measurements of $\Omega$ and $\Lambda$ from 42 high redshift supernovae'',
Astrophys. J. \textbf{517}, 565-586 (1999)
doi:10.1086/307221
[arXiv:astro-ph/9812133 [astro-ph]].	

\bibitem{Copeland06}
E.~J.~Copeland, M.~Sami and S.~Tsujikawa,
``Dynamics of dark energy,''
Int. J. Mod. Phys. D \textbf{15}, 1753-1936 (2006)
doi:10.1142/S021827180600942X
[arXiv:hep-th/0603057 [hep-th]].
	
\bibitem{Cai10}
Y.~F.~Cai, E.~N.~Saridakis, M.~R.~Setare and J.~Q.~Xia,
``Quintom Cosmology: Theoretical implications and observations,''
Phys. Rept. \textbf{493}, 1-60 (2010)
doi:10.1016/j.physrep.2010.04.001
[arXiv:0909.2776 [hep-th]].

\bibitem{Bamba12}
K.~Bamba, S.~Capozziello, S.~Nojiri and S.~D.~Odintsov,
``Dark energy cosmology: the equivalent description via different theoretical models and cosmography tests,''
Astrophys. Space Sci. \textbf{342}, 155-228 (2012)
doi:10.1007/s10509-012-1181-8
[arXiv:1205.3421 [gr-qc]].

\bibitem{Nojiri10}
S.~Nojiri and S.~D.~Odintsov,
``Unified cosmic history in modified gravity: from F(R) theory to Lorentz non-invariant models,''
Phys. Rept. \textbf{505}, 59-144 (2011)
doi:10.1016/j.physrep.2011.04.001
[arXiv:1011.0544 [gr-qc]].

\bibitem{Capozziello11}
S.~Capozziello and M.~De Laurentis,
``Extended Theories of Gravity,''
Phys. Rept. \textbf{509}, 167-321 (2011)
doi:10.1016/j.physrep.2011.09.003
[arXiv:1108.6266 [gr-qc]].

\bibitem{Cai16}
Y.~F.~Cai, S.~Capozziello, M.~De Laurentis and E.~N.~Saridakis,
``f(T) teleparallel gravity and cosmology,''
Rept. Prog. Phys. \textbf{79}, no.10, 106901 (2016)
doi:10.1088/0034-4885/79/10/106901
[arXiv:1511.07586 [gr-qc]].

\bibitem{CANTATA21}
E.~N.~Saridakis \textit{et al.} [CANTATA],
``Modified Gravity and Cosmology: An Update by the CANTATA Network,''
[arXiv:2105.12582 [gr-qc]].

\bibitem{Hooft93}
G.~'t Hooft,
``Dimensional reduction in quantum gravity,''
Conf. Proc. C \textbf{930308}, 284-296 (1993)
[arXiv:gr-qc/9310026 [gr-qc]].

\bibitem{Susskind95}
L.~Susskind,
``The World as a hologram,''
J. Math. Phys. \textbf{36}, 6377-6396 (1995)
doi:10.1063/1.531249
[arXiv:hep-th/9409089 [hep-th]].

\bibitem{Bousso02}
R.~Bousso,
``The Holographic principle,''
Rev. Mod. Phys. \textbf{74}, 825-874 (2002)
doi:10.1103/RevModPhys.74.825
[arXiv:hep-th/0203101 [hep-th]].

\bibitem{Bekenstein73}
J.~D.~Bekenstein,
``Black holes and entropy,''
Phys. Rev. D \textbf{7}, 2333-2346 (1973)
doi:10.1103/PhysRevD.7.2333.

\bibitem{Hawking75}
S.~W.~Hawking,
``Particle Creation by Black Holes,''
Commun. Math. Phys. \textbf{43}, 199-220 (1975)
[erratum: Commun. Math. Phys. \textbf{46}, 206 (1976)]
doi:10.1007/BF02345020.

\bibitem{Li04}
M.~Li,
``A Model of holographic dark energy,''
Phys. Lett. B \textbf{603}, 1 (2004)
doi:10.1016/j.physletb.2004.10.014
[arXiv:hep-th/0403127 [hep-th]].

\bibitem{SWang17}
S.~Wang, Y.~Wang and M.~Li,
''Holographic Dark Energy,''
Phys. Rept. \textbf{696}, 1-57 (2017)
doi:10.1016/j.physrep.2017.06.003
[arXiv:1612.00345 [astro-ph.CO]].

\bibitem{Horvat04}
R.~Horvat,
``Holography and variable cosmological constant,''
Phys. Rev. D \textbf{70}, 087301 (2004)
doi:10.1103/PhysRevD.70.087301
[arXiv:astro-ph/0404204 [astro-ph]].

\bibitem{Pavon05}
D.~Pavon and W.~Zimdahl,
``Holographic dark energy and cosmic coincidence,''
Phys. Lett. B \textbf{628}, 206-210 (2005)
doi:10.1016/j.physletb.2005.08.134
[arXiv:gr-qc/0505020 [gr-qc]].

\bibitem{Nojiri06}
S.~Nojiri and S.~D.~Odintsov,
``Unifying phantom inflation with late-time acceleration: Scalar phantom-non-phantom transition model and generalized holographic dark energy,''
Gen. Rel. Grav. \textbf{38}, 1285-1304 (2006)
doi:10.1007/s10714-006-0301-6
[arXiv:hep-th/0506212 [hep-th]].

\bibitem{BWang06}
B.~Wang, C.~Y.~Lin and E.~Abdalla,
``Constraints on the interacting holographic dark energy model,''
Phys. Lett. B \textbf{637}, 357-361 (2006)
doi:10.1016/j.physletb.2006.04.009
[arXiv:hep-th/0509107 [hep-th]].

\bibitem{Setare09Saridakis}
M.~R.~Setare and E.~N.~Saridakis,
``Non-minimally coupled canonical, phantom and quintom models of holographic dark energy,''
Phys. Lett. B \textbf{671}, 331-338 (2009)
doi:10.1016/j.physletb.2008.12.026
[arXiv:0810.0645 [hep-th]].

\bibitem{Cai07}
R.~G.~Cai,
``A Dark Energy Model Characterized by the Age of the Universe,''
Phys. Lett. B \textbf{657}, 228-231 (2007)
doi:10.1016/j.physletb.2007.09.061
[arXiv:0707.4049 [hep-th]].

\bibitem{Jawad16}
A.~Jawad, N.~Azhar and S.~Rani,
``Entropy corrected holographic dark energy models in modified gravity,''
Int. J. Mod. Phys. D \textbf{26}, no.04, 1750040 (2016)
doi:10.1142/S0218271817500407.

\bibitem{Pasqua16}
A.~Pasqua, S.~Chattopadhyay and R.~Myrzakulov,
``Power-law entropy-corrected holographic dark energy in Ho\v{r}ava-Lifshitz cosmology with Granda-Oliveros cut-off,''
Eur. Phys. J. Plus \textbf{131}, no.11, 408 (2016)
doi:10.1140/epjp/i2016-16408-8
[arXiv:1511.00611 [gr-qc]].

\bibitem{Pourhassan18}
B.~Pourhassan, A.~Bonilla, M.~Faizal and E.~M.~C.~Abreu,
``Holographic Dark Energy from Fluid/Gravity Duality Constraint by Cosmological Observations,''
Phys. Dark Univ. \textbf{20}, 41-48 (2018)
doi:10.1016/j.dark.2018.02.006
[arXiv:1704.03281 [hep-th]].

\bibitem{Nojiri17}
S.~Nojiri and S.~D.~Odintsov,
``Covariant Generalized Holographic Dark Energy and Accelerating Universe,''
Eur. Phys. J. C \textbf{77}, no.8, 528 (2017)
doi:10.1140/epjc/s10052-017-5097-x
[arXiv:1703.06372 [hep-th]].

\bibitem{Zhang05}
X.~Zhang and F.~Q.~Wu,
``Constraints on holographic dark energy from Type Ia supernova observations,''
Phys. Rev. D \textbf{72}, 043524 (2005)
doi:10.1103/PhysRevD.72.043524
[arXiv:astro-ph/0506310 [astro-ph]].

\bibitem{Li09}
M.~Li, X.~D.~Li, S.~Wang and X.~Zhang,
``Holographic dark energy models: A comparison from the latest observational data,''
JCAP \textbf{06}, 036 (2009)
doi:10.1088/1475-7516/2009/06/036
[arXiv:0904.0928 [astro-ph.CO]].

\bibitem{Feng07}
C.~Feng, B.~Wang, Y.~Gong and R.~K.~Su,
``Testing the viability of the interacting holographic dark energy model by using combined observational constraints,''
JCAP \textbf{09}, 005 (2007)
doi:10.1088/1475-7516/2007/09/005
[arXiv:0706.4033 [astro-ph]].

\bibitem{Zhang09}
X.~Zhang,
``Holographic Ricci dark energy: Current observational constraints, quintom feature, and the reconstruction of scalar-field dark energy,''
Phys. Rev. D \textbf{79}, 103509 (2009)
doi:10.1103/PhysRevD.79.103509
[arXiv:0901.2262 [astro-ph.CO]].


\bibitem{Lu10}
J.~Lu, E.~N.~Saridakis, M.~R.~Setare and L.~Xu,
``Observational constraints on holographic dark energy with varying gravitational constant,''
JCAP \textbf{03}, 031 (2010)
doi:10.1088/1475-7516/2010/03/031
[arXiv:0912.0923 [astro-ph.CO]].

\bibitem{Tsallis12}
C.~Tsallis and L.~J.~L.~Cirto,
``Black hole thermodynamical entropy,''
Eur. Phys. J. C \textbf{73}, 2487 (2013)
doi:10.1140/epjc/s10052-013-2487-6
[arXiv:1202.2154 [cond-mat.stat-mech]].

\bibitem{Tavayef:2018xwx}
M.~Tavayef, A.~Sheykhi, K.~Bamba and H.~Moradpour,
``Tsallis Holographic Dark Energy,''
Phys. Lett. B \textbf{781}, 195-200 (2018)
doi:10.1016/j.physletb.2018.04.001.

\bibitem{Kaniadakis02}
G.~Kaniadakis,
``Statistical mechanics in the context of special relativity,''
Phys. Rev. E \textbf{66}, 056125 (2002)
doi:10.1103/PhysRevE.66.056125
[arXiv:cond-mat/0210467 [cond-mat.stat-mech]].

\bibitem{Kaniadakis05}
G.~Kaniadakis,
``Statistical mechanics in the context of special relativity. II.,''
Phys. Rev. E \textbf{72}, 036108 (2005)
doi:10.1103/PhysRevE.72.036108
[arXiv:cond-mat/0507311 [cond-mat]].

\bibitem{Drepanou21}
N.~Drepanou, A.~Lymperis, E.~N.~Saridakis and K.~Yesmakhanova,
``Kaniadakis holographic dark energy,''
[arXiv:2109.09181 [gr-qc]].

\bibitem{Sharma21}
U.~K.~Sharma, V.~C.~Dubey, A.~H.~Ziaie and H.~Moradpour,
``Kaniadakis Holographic Dark Energy in non-flat Universe,''
[arXiv:2106.08139 [physics.gen-ph]].




\bibitem{Moradpour20}
H.~Moradpour, A.~H.~Ziaie and M.~Kord Zangeneh,
``Generalized entropies and corresponding holographic dark energy models,''
Eur. Phys. J. C \textbf{80}, no.8, 732 (2020)
doi:10.1140/epjc/s10052-020-8307-x
[arXiv:2005.06271 [gr-qc]].

\bibitem{Jawad21}
A.~Jawad and A.~M.~Sultan,
``Cosmic Consequences of Kaniadakis and Generalized Tsallis Holographic Dark Energy Models in the Fractal Universe,''
Adv. High Energy Phys. \textbf{2021}, 5519028 (2021)
doi:10.1155/2021/5519028

\bibitem{Nisha21}
N.~M.~Ali, Pankaj, U.~K.~Sharma, S.~K.~P and S.~Srivastava,
``New Tsallis holographic dark energy with apparent horizon as IR-cutoff in non-flat Universe,''
[arXiv:2110.07021 [physics.gen-ph]].




\bibitem{Srednicki93}
M.~Srednicki,
``Entropy and area,''
Phys. Rev. Lett. \textbf{71}, 666-669 (1993)
doi:10.1103/PhysRevLett.71.666
[arXiv:hep-th/9303048 [hep-th]].

\bibitem{Das07}
S.~Das and S.~Shankaranarayanan,
``Where are the black hole entropy degrees of freedom?,''
Class. Quant. Grav. \textbf{24}, 5299-5306 (2007)
doi:10.1088/0264-9381/24/20/022
[arXiv:gr-qc/0703082 [gr-qc]].

\bibitem{Pavon20}
D.~Pavon,
``On the degrees of freedom of a black hole,''
[arXiv:2001.05716 [gr-qc]].

\bibitem{Abreu18}
E.~M.~C.~Abreu, J.~A.~Neto, A.~C.~R.~Mendes, A.~Bonilla and R.~M.~de Paula,
``Tsallis' entropy, modified Newtonian accelerations and the Tully-Fisher relation,''
EPL \textbf{124}, no.3, 30005 (2018)
doi:10.1209/0295-5075/124/30005
[arXiv:1804.06723 [hep-th]].

\bibitem{Masi05}
Marco~Masi
``A step beyond Tsallis and Rényi entropies,''
Physics Letters A \textbf{338}, 217 (2005)
doi:10.1016/j.physleta.2005.01.094
[arXiv:cond-mat/0505107].

\bibitem{Hsu04}
S.~D.~H.~Hsu,
\textit{Entropy bounds and dark energy},
Phys. Lett. B \textbf{594}, 13-16 (2004)
doi:10.1016/j.physletb.2004.05.020
[arXiv:hep-th/0403052 [hep-th]].

\bibitem{Zhang07}
X.~Zhang,
\textit{Reconstructing holographic quintessence},
Phys. Lett. B \textbf{648}, 1-7 (2007)
doi:10.1016/j.physletb.2007.02.069
[arXiv:astro-ph/0604484 [astro-ph]].


\end{thebibliography}
\end{document}